\documentclass[a4paper,fleqn,usenatbib]{mnras}

\usepackage[T1]{fontenc}

\usepackage{newtxtext,newtxmath}

\usepackage{graphicx}	
\usepackage{amsmath}	
\usepackage{amssymb}	

\usepackage{xspace}

\newcommand{\fref}[1]{Fig.~\ref{#1}\xspace}
\newcommand{\eref}[1]{eq.~\eqref{#1}\xspace}
\newcommand{\sref}[1]{Section~\ref{#1}\xspace}

\newcommand{\msun}{\ensuremath{\, \text{M}_{\odot}}\xspace}
\newcommand{\kel}{\ensuremath{\, \text{K}}\xspace}

\newcommand{\ttau}{\ensuremath{T(\tau)}\xspace}
\newcommand{\ttaurel}{\ttau relation\xspace}
\newcommand{\taueff}{\ensuremath{\tau_{\textup{eff}}}\xspace}
\newcommand{\tausurf}{\ensuremath{\tau_{\textup{surf}}}\xspace}
\newcommand{\taufit}{\ensuremath{\tau_{\textup{fit}}}\xspace}
\newcommand{\hp}{\ensuremath{H_{\textup{p}}}\xspace}
\newcommand{\mlt}{\ensuremath{\alpha_{\textup{\tiny MLT}}}\xspace}
\newcommand{\mlts}{\ensuremath{\alpha_{\textup{\tiny MLT, }\odot}}\xspace}
\newcommand{\mltg}{\ensuremath{\alpha_{\textup{{\tiny MLT}, grid}}}\xspace}
\newcommand{\mltgs}{\ensuremath{\alpha_{\textup{{\tiny MLT}, grid}, \odot}}\xspace}
\newcommand{\logg}{\ensuremath{\log g}\xspace}
\newcommand{\teff}{\ensuremath{T_{\textup{eff}}}\xspace}
\newcommand{\radgrad}{\ensuremath{\nabla_{\textup{rad}}}\xspace}
\newcommand{\radgradt}{\ensuremath{\widetilde{\nabla}_{\textup{rad}}}\xspace}
\newcommand{\dq}{\ensuremath{q\;\!'\,\!(\tau)}\xspace}
\newcommand{\dqq}{\ensuremath{q'(\tau)}\xspace}
\newcommand{\gar}{\textsc{garstec}\xspace}
\newcommand{\mes}{\textsc{mesa}\xspace}
\newcommand{\adi}{\textsc{adipls}\xspace}
\newcommand{\kmlt}{\ensuremath{K_{\textup{\tiny MLT}}}\xspace}

\title[Stellar models with calibrated 3D results]{Stellar models with calibrated
  convection and temperature stratification from 3D hydrodynamics simulations}

\author[J. R. Mosumgaard et al.]{%
Jakob R\o{}rsted Mosumgaard,$^{1,2}$\thanks{E-mail: jakob@phys.au.dk}
Warrick H.~Ball,$^{3,1}$
V\'{i}ctor Silva Aguirre,$^{1}$\newauthor
Achim Weiss$^{2}$, and J{\o}rgen Christensen-Dalsgaard$^{1}$
\\
$^{1}$ Stellar Astrophysics Centre (SAC), Department of Physics and
Astronomy, Aarhus University, Ny Munkegade 120, DK-8000 Aarhus C, Denmark\\
$^{2}$ Max-Planck-Institut f\"{u}r Astrophysik, Karl-Schwarzschild-Str. 1, D-85748 Garching, Germany\\
$^{3}$ School of Physics \& Astronomy, University of Birmingham, Edgbaston, Birmingham B15 2TT, UK\\
}

\date{Accepted 2018 May 31. Received 2018 May 16; in original form 2018 March 27}

\pubyear{2018}

\begin{document}
\label{firstpage}
\pagerange{\pageref{firstpage}--\pageref{lastpage}}
\maketitle

\begin{abstract}
  Stellar evolution codes play a major role in present-day astrophysics, yet
  they share common simplifications related to the outer layers of stars. We
  seek to improve on this by the use of results from realistic and highly
  detailed 3D hydrodynamics simulations of stellar convection. We implement a
  temperature stratification extracted directly from the 3D simulations into two
  stellar evolution codes to replace the simplified atmosphere normally used.
  Our implementation also contains a non-constant mixing-length parameter, which
  varies as a function of the stellar surface gravity and temperature -- also
  derived from the 3D simulations. We give a detailed account of our fully
  consistent implementation and compare to earlier works, and also provide a
  freely available \mes-module. The evolution of low-mass stars with different
  masses is investigated, and we present for the first time an asteroseismic
  analysis of a standard solar model utilising calibrated convection and
  temperature stratification from 3D simulations. We show that the inclusion of
  3D results have an almost insignificant impact on the evolution and structure
  of stellar models -- the largest effect are changes in effective temperature
  of order $30\kel$ seen in the pre-main sequence and in the red-giant branch.
  However, this work provides the first step for producing self-consistent
  evolutionary calculations using fully incorporated 3D atmospheres from
  on-the-fly interpolation in grids of simulations.
\end{abstract}






\section{Introduction}
\label{sec:intro}

Understanding stellar structure and evolution is one of the key ingredients in
astrophysics. One of the primary tools for doing so is comparing numerical
calculations of stellar structure with observations and analysing changes in
time using a stellar evolution code. These are one-dimensional (1D) numerical
models, which have been tested and developed through the years; as a result they
are very efficient and highly optimised. However, in several aspects they are
also simplified and can be improved.

The first problem we address in this paper is related to the outer boundary
conditions of the models, which are required to solve the stellar equations.
Traditionally they are established by integration of a simplified temperature
stratification as a function of optical depth -- a so-called \ttaurel -- which
combined with hydrostatic equilibrium provides the pressure for the outermost
point in the model. The most commonly used is the analytical grey Eddington
atmosphere \citep[e.g.][]{Kippenhahn2012}; other popular choices are
semi-empirical relations derived from the Sun
\citep[][]{KrishnaSwamy1966,Vernazza1981}. The \ttaurel (also commonly known as
the atmosphere) employed in the model has a non-negligible impact on the
structure and evolution \citep[see, e.g.][]{Salaris2002,Tanner2014}.

Secondly we address one of the most fundamental problems in stellar physics: the
treatment of stellar convection. The most adopted method for parametrising this
process in stellar evolution calculations is the formalism from
\citet{Bohm-Vitense1958} known as the mixing-length theory (MLT). The basic idea
in MLT is to model convection as rising and falling elements, which move in a
background medium. The background is the mean stratification where pressure
equilibrium is assumed as well as symmetry between the up- and downflows. A
convective element moves a certain distance $\Lambda$ before dissolving/mixing
instantaneously into the ambient medium. This distance (called the mixing
length) is assumed proportional to the local pressure scale height \hp as
$\Lambda = \mlt \hp$, where \mlt is the central free parameter of the theory:
the mixing-length parameter.

Usually the value of \mlt is constrained by the Sun. However, in several recent
studies on the detailed modelling of red-giant stars, \mlt has been highlighted
as a focus point. For instance \citet{Li2018} reported the need of modifying
\mlt in the RGB by using eclipsing binaries from \emph{Kepler} as calibrators.
\Citet{Tayar2017} found similar problems based on several thousand red giants --
although this claim has been disputed in the very recent work by
\citet{Salaris2018x}. \Citet{Hjorringgaard2016} have reported on discrepancies
between constraints from different independent methods when determining the
parameters of the very well-studied red giant HD 185351. Moreover, the problem
of an unconstrained \mlt is not unique to red giants: \citet{White2017} showed
that their parameter determinations using \mes changed significantly when \mlt
was free to vary compared to a fixed solar value.

Another very different approach to superadiabatic convection in the outer layers
of stars is to make a full three-dimensional (3D) simulation using
radiation-coupled hydrodynamics (RHD). These simulations cover both the
radiative outer atmosphere as well as the superadiabatic convective
sub-photospheric layers, reaching the quasi-adiabatic deeper convective layers.
No parametric theory is needed to generate the convection; only fundamental
physics. These simulations are inherently more realistic, and have become very
popular in recent years. The 3D simulations have produced impressive results --
e.g. being able to reproduce the observed solar granulation from first
principles \citep{Stein1998} -- and have altered our understanding of stellar
convection \citep[e.g.][]{Stein1989,Nordlund2009}. A major drawback of this kind
of simulations is that they are computationally very expensive. Moreover, the
simulations are designed to study the stellar granulation, which operates on
very short timescales compared to nuclear processes in the star. Thus, these
simulations cannot be directly utilised to calculate stellar evolution models,
as they are not able to follow the timescales required.

In this work, we seek to remedy some of the issues in stellar evolution models
by the use of results from the realistic and highly detailed 3D RHD simulations
of stellar convection. The ultimate goal is to employ full structures from the
3D simulations on-the-fly in stellar evolutionary calculations, thus producing
the next generation of stellar models. This paper -- which continues the work of
\citet{Mosumgaard2017x} -- is a major step in the pursuit of this.
%

The paper is organised as follows. In the next section, a description of the 3D
simulations and the data used is given. We present our implementation in some
detail in \sref{sec:implement} and the impact of using it in evolutionary and
asteroseismic calculations in \sref{sec:results}. In \sref{sec:discussion} we
present a discussion, where we also compare our implementation with the previous
work by \citet{Salaris2015} -- also elaborated in Appendix~\ref{app:sc15} -- and
finally a conclusion in \sref{sec:conclusion}. In Appendix~\ref{app:mesa} we
review the technical details of our \mes-module, which we make freely available.

\section{3D simulations of stellar atmospheres}
\label{sec:atmos}

\begin{figure}
  \centering
  \includegraphics[width=\columnwidth]{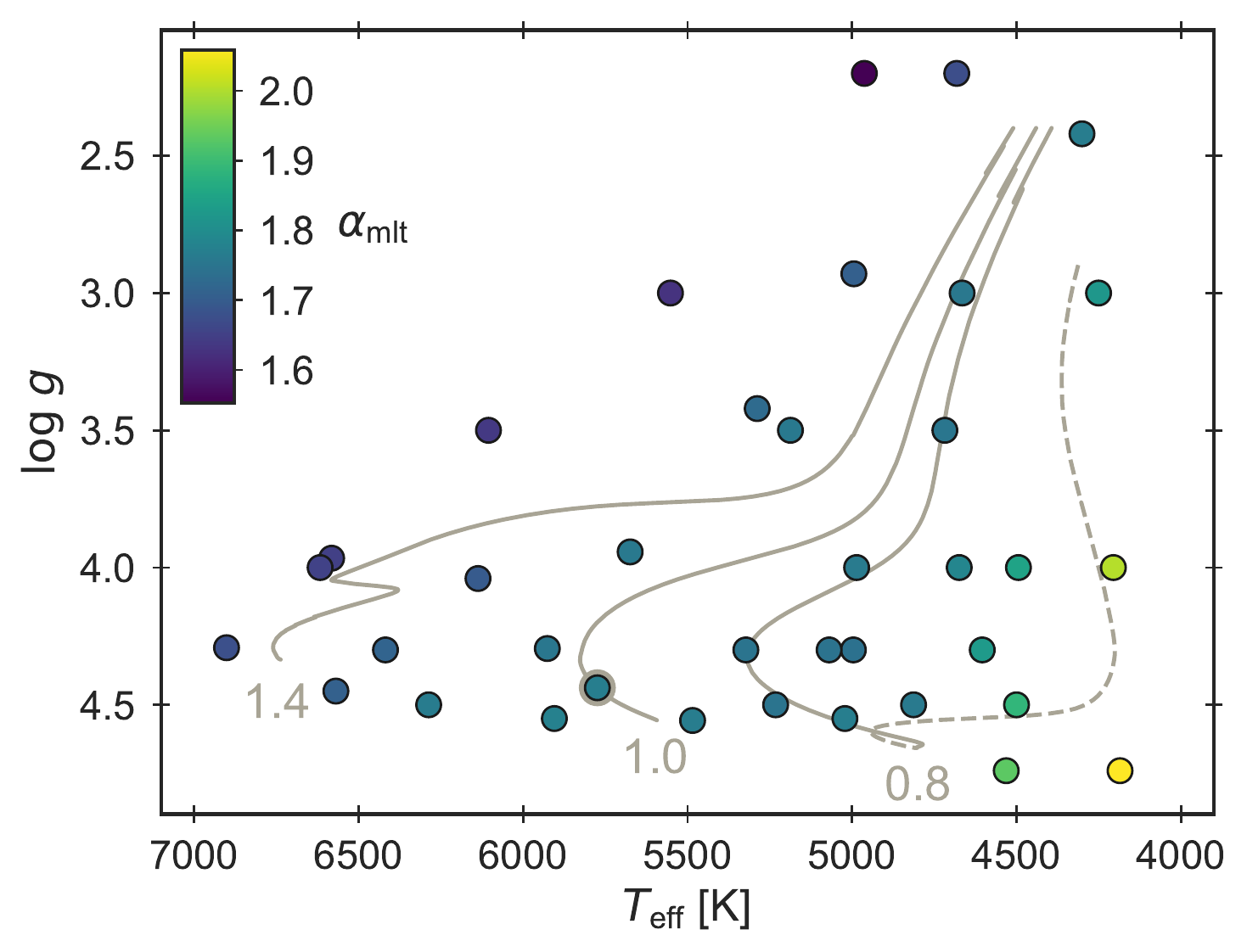}
  \caption{A Kiel diagram, in terms of effective temperature \teff and logarithm
    of surface gravity $g$, showing the grid of 3D simulations at solar
    metallicity coloured to show the corresponding value of the calibrated
    mixing-length parameter \mlt{} from \citet{Trampedach2014b}. Evolutionary
    tracks from \gar for stars of different masses (in units of the mass of the
    Sun) are shown for reference -- the pre-main sequence for one of the tracks
    are shown with dashes.}
  \label{fig:grid}
\end{figure}

We have used data from the grid of 37 atmospheric 3D simulations of stellar
convection by \citet{Trampedach2013,Trampedach2014a,Trampedach2014b}. All of the
simulations are computed with solar metallicity and the coverage in a Kiel
diagram (effective temperature \teff and surface gravity \logg) is shown in
\fref{fig:grid}. It should be noted that the simulations were calculated using a
non-standard solar composition based on \citet[AG89]{Anders1989}; the exact
composition -- which is actually very close to \citet[GN93]{Grevesse1993x} -- is
given in \citet[Table~1]{Trampedach2013}.
%

\subsection{Patched models}
\label{sec:atmos_patch}

The most common synergy between 3D convection simulations and typical stellar
evolution models are the so-called \emph{patched models}. The principle in
producing a patched model is to peel off the surface layers of a stellar
structure model and replace it with data from a 3D simulation averaged over time
and the horizontal directions. Great care must be taken in ensuring a correct
and compatible patching between the two different parts -- e.g. by using similar
microphysics and boundary conditions \citep{Trampedach2014a}. Several authors
have utilised this technique and it has primarily been applied in helio- and
asteroseismology to investigate the so-called \emph{asteroseismic surface
  effect} \citep{Brown1984,Christensen-Dalsgaard1988}, as the procedure
naturally alters the outer regions in the models.

The technique was originally applied to the Sun by \citet{Rosenthal1999}, and
later by \citet{Piau2014} and \citet{Magic2016}. It has also been used in
analyses across different stellar parameters by e.g. \citet{Sonoi2015},
\citet{Ball2016}, and \citet{Trampedach2017}. The advantage of this method is
that the poorly modelled surface layers of the stellar model are fully replaced
by the sophisticated 3D structure, which significantly diminishes the surface
effect. However, the method comes with two primary disadvantages.

A first clear issue is that the simulations only sample discrete points in the
HR diagram. Thus, in order to patch, the interior model must correspond to a
specific 3D simulation, making it impractical when modelling ``real'' stars.
This part of the problem has been treated by \citet{Joergensen2017}, who
developed a technique for interpolating between the 3D simulations in a grid.
The work utilized the interpolation scheme for producing the patched models with
arbitrary physical parameters, and applied the method to several stars observed
by the \emph{Kepler} space mission \citep{Borucki2010x}. In fact
\citet{Joergensen2017} calculated their pre-patching evolution using the
\gar-implementation described in this paper, before patching with the
corresponding 3D simulations, to ensure the highest level of consistency.

Secondly, it is not possible to do time evolution with the patched models. It is
a purely static approach: The 1D model is evolved using a regular boundary
condition until a certain point, where it is then patched to a corresponding 3D
simulation. Stellar evolution is the main focus of the dynamic implementation we
present in the following.

\subsection{Condensed simulations}
\label{sec:atmos_condensed}

As a first step to overcoming some of the problems,
\citet{Trampedach2014a,Trampedach2014b} developed a method for extracting
information from the 3D simulations targeted at the two issues mentioned above.

Firstly, they have used 1D envelope models to calibrate the mixing-length
parameter, \mlt, for each simulation. The basic principle behind the calibration
is similar to the patching procedure described before, but using envelope models
instead of full structure models -- and of course varying \mlt as a part of the
process. The authors went to great length to ensure compatible (micro)physics
between their 1D and 3D models. The exact procedure is described by
\citet{Trampedach2014b} and the resulting values are visualised in
\fref{fig:grid}. Stellar evolution tracks of different masses are also shown in
the figure to display the coverage of the grid. A similar work was done using 2D
simulations by \citet{Ludwig1999}, who found similar trends with stellar
parameters.

Secondly, the authors devised a way of distilling the averaged temperature
stratification from the full 3D simulations. They have taken great care in the
treatment of convective effects and enforcing radiative equilibrium; a detailed
description is given by \citet{Trampedach2014a}. The output is given in the form
of a \emph{generalised} Hopf function $q(\tau)$ given as
\begin{align}
  \label{eq:generalhopf}
  \frac{4}{3} \left( \frac{T}{\teff} \right)^{4} = q(\tau) + \tau \; ,
\end{align}
where $\tau$ is a Rosseland mean optical depth \citep[eq.~9]{Trampedach2014a}.
It is important to note that these extracted \ttaurel{}s were also used in
\citet{Trampedach2014b} to perform the \mlt-calibration mentioned above in order
to ensure maximum consistency.

The results provided by \citet{Trampedach2014a,Trampedach2014b} -- i.e $q(\tau)$
and calibrated \mlt for all simulations -- are stored as a function of \teff and
\logg. The main advantage of the described approach is that the extracted
quantities behave smoothly with varying \teff and \logg, making it feasible to
directly interpolate between the simulations. Thus, it is possible to include
the 3D effects during the evolution of stars within the full range of the grid
of simulations.

\section{Implementation}
\label{sec:implement}

We have consistently implemented the use of information from 3D atmospheric
simulations from \citet{Trampedach2014a,Trampedach2014b} into the Garching
Stellar Evolution Code \citep[\gar,][]{Weiss2008} and Modules for Experiments in
Stellar Astrophysics \citep[\mes,][]{Paxton2011x,Paxton2013x,Paxton2015x}. The
basic principle of our implementation is that in each iteration of the stellar
evolution code, the current \teff and \logg of the model are used to obtain
corresponding information from the 3D simulations. More specifically we extract
the \mlt and corresponding $q(\tau)$ -- which can be transformed to a \ttaurel{}
(see below) -- from a table as a function of \teff and \logg. To interpolate in
the irregular grid of simulations, we use the routine supplied by
\citet{Trampedach2014a,Trampedach2014b}, which is based on linear interpolation
between the nodes in a Thiessen triangulation \citep{Cline1984}.

\subsection{Mixing-length parameter}
\label{sec:implement_alpha}

The first change to the stellar structure model involves the mixing-length
parameter, which is vital for the treatment of convection. Instead of a constant
value we use the variable 3D-calibrated mixing-length parameter, $\mlt(\teff,
\logg)$, throughout the model.

We are not taking the value directly as provided by the interpolation in the
table, but introduce a scaling factor as recommended by \citet{Ludwig1999} and
\citet{Trampedach2014b}. The actual \mlt used in the code is given by
\begin{equation}
  \label{eq:mlt}
  \mlt(\teff, \logg) = \kmlt \cdot \mltg(\teff, \logg) \; ,
\end{equation}
where \kmlt is the scaling factor and \mltg is the value returned from the
interpolation routine.

The scaling factor must be determined from a solar calibration with the 3D
\ttaurel (see below) and variable mixing-length parameter activated. This
scaling ensures that the solar model is calibrated to the correct radius with
the variable \mlt.

\subsection[T-tau relation]{\ttaurel{}}
\label{sec:implement_ttau}

The other aspect of our implementation is related to the use of a realistic
\ttaurel derived from the 3D simulations. As mentioned above, the interpolation
routine supplies the current stratification (updated in each iteration) as a
function of optical depth, $\tau$, in terms of the generalised Hopf function,
$q(\tau)$.

The changes made in \gar and \mes are nearly identical. We provide a detailed
account of the \gar implementation and at the end highlight how the
modifications in the two codes differ. We will split the description in two (one
for each part of a \gar structure model): atmosphere and interior.

\subsubsection{Atmospheric part}
\label{sec:implement_ttau_atmos}

The atmospheric part of the model is used to provide the outer boundary
conditions for the equations of stellar structure and evolution. In the standard
setting, the code uses an Eddington grey atmosphere, which is anchored to the
interior model at an optical depth of $\tau = 2/3$.

In our implementations, we have altered the atmospheric module to obtain the
temperature stratification from the generalised Hopf function $q(\tau)$
extracted from the 3D simulations. The new temperature structure at a given
optical depth can be obtained from \eref{eq:generalhopf} as
\begin{align}
  \label{eq:hopf}
  T(\tau) = \left( \frac{3}{4} \right)^{1/4} \left[ q(\tau) + \tau \right]^{1/4}
  \teff \; ,
\end{align}
under the assumption of radiative equilibrium.

In \gar, the stratification is integrated inwards from $\tausurf = 10^{-4}$ to
obtain the pressure at the bottom of the atmosphere. This point is the fitting
point or transition point to the interior model -- commonly referred to as the
photosphere -- which is defined to have the temperature $T = \teff$. As opposed
to the Eddington grey case, the 3D \ttaurel{} from the generalised Hopf function
is not anchored at $\taufit = 2/3$, or any other constant value of $\tau$.
Instead, we require that the temperature of the outermost point in the interior
model matches that of the bottom point in the atmosphere.
Thus the fitting point $\taufit = \taueff$ is defined at the value of $\tau$
where $T = \teff$, or
\begin{align}
  q(\taueff) + \taueff = 4/3 \; .
\end{align}
This value is not constant, which means that in each iteration of the code
before the actual integration, the point \taueff is determined from the current
3D \ttaurel{} with interpolation. Usually $\taueff \simeq 0.5$ is obtained.

It is very important to choose the correct optical depth of the photosphere of
the stellar structure model; otherwise \teff and \logg of the model will not
actually correspond to the output from the 3D interpolation. In other words,
selecting a constant optical depth (e.g. $\tau = 2/3$) for the photosphere of
the model -- hence defining \teff at this point -- will clearly lead to small
inconsistencies if used with a \ttaurel where \teff corresponds to a different
optical depth.

\subsubsection{Interior part}
\label{sec:implement_ttau_interior}

Besides the above-mentioned change to a variable mixing-length parameter, our
implementation directly modifies the interior part of the model as well.

As explained earlier, the $q(\tau)$ is extracted from the 3D simulations under
the assumption of radiative equilibrium, which is also assumed when using
\eref{eq:hopf} for the temperature structure. Nonetheless, we want to properly
take convection into account below \taueff to recover the correct stratification
from the 3D simulations. According to \citet[eq.~35]{Trampedach2014a}, the
radiative temperature gradient, $\radgrad\equiv(\partial\ln T/\partial\ln
p)_{\textup{rad}}$, therefore needs to be altered. The corrected gradient is
calculated as
\begin{equation}
  \label{eq:radgrad}
  \radgrad = \radgradt \cdot [1 + \dq] \; ,
\end{equation}
where \radgradt is the usual expression for the radiative gradient based on the
diffusion approximation and \dqq is the derivative of the Hopf function with
respect to $\tau$ (which in \gar is determined directly from an Akima spline
interpolation \citep{Akima1970,Akima1991} of $q(\tau)$ on a fine
mesh)\footnote{We use this method because the derivative \dqq is changing
  rapidly in the region around $\tau = \taueff \simeq 0.5$.}.

The radiative gradient is corrected before performing the actual MLT
calculation. Hence, the modified \radgrad is used as input instead of the usual
\radgradt, such that the resulting temperature gradient $\nabla$ is properly
corrected for 3D effects and does not need to be modified further.

The altered radiative gradient is used until an optical depth of $\tau = 10$ is
reached, since the correction factor goes to $0$; \dqq is always below $10^{-4}$
(and usually below $10^{-5}$) from this point inwards. Our implementation is
fully flexible with respect to this lower point and the effect of changing it to
a higher value (the table extends down to $\tau = 100$) is completely negligible
as the corrections in this region are minute.

\begin{figure}
  \centering
  \includegraphics[width=0.7\columnwidth]{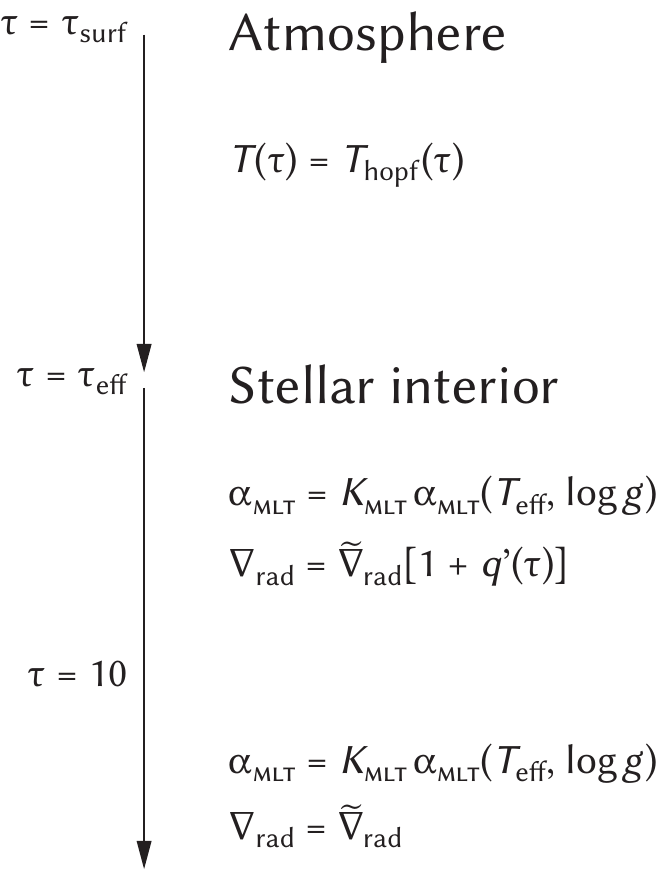}
  \caption{The changes to a \gar stellar structure model as a result of our
    implementation. In the radiative atmosphere, a different temperature
    stratification, given by \eref{eq:hopf}, is used and the transition point is
    altered. In the outer, convective parts of the interior model, the radiative
    gradient is modified according to \eref{eq:radgrad}. Everywhere in the
    interior of the star, the calibrated $\mlt(\teff,\logg)$ from \eref{eq:mlt}
    is used. In \mes, the \emph{stellar interior} extends all the way to
    \tausurf, but the photospheric quantities are determined at $\tau =
    \taueff$.}
  \label{fig:model}
\end{figure}

A full schematic overview and summary of how our implementation changes a
stellar model (both atmospheric and interior part) can be seen in
\fref{fig:model}.

\subsubsection{MESA}
\label{sec:implement_ttau_mesa}
The \mes implementation incorporates the stellar atmosphere as part of the
interior model by placing the outermost meshpoint at an optical depth $\tausurf
= 2 \times 10^{-4} \taueff \approx 10^{-4}$. However, all photospheric
quantities are determined at \taueff by interpolation. The correct temperature
stratification for the 3D \ttaurel is obtained by the same method that \gar uses
in the interior below the photosphere (\eref{eq:radgrad}). The only distinction
is the interpolation in $\tau$, which is performed using the one-dimensional
monotone cubic piecewise interpolation routines distributed with \mes
\citep{Huynh1993, Suresh1997}. The radiative gradient is modified using the
so-called `porosity factor' (see Appendix~\ref{app:rad_grad}).

The new outermost meshpoint requires a new boundary conditions, which we choose
to be equivalent to an Eddington grey atmosphere evaluated at the optical depth
of the point (details given in the Appendix~\ref{app:mesa}). In effect, these
are the same as the boundary conditions \mes uses when integrating an atmosphere
down to the photosphere.
%

\section{Stellar Evolution with 3D results}
\label{sec:results}

We have taken much care to adopt as similar as possible microphysics in the
stellar evolution code as in the 3D simulations in order to produce a consistent
model.

The envelope models which \citet{Trampedach2014b} utilised to perform the 3D
\mlt-calibration were calculated with the MLT treatment from
\citet{Bohm-Vitense1958}. To include the calibrated \mlt values, we also use a
standard mixing-length theory of convection in our stellar evolution
calculations. Strictly speaking, the \mlt calibration is only valid for
precisely the same MLT implementation. But it is important to note, that even
different MLT-flavors will yield the same temperature evolution, if \mlt is
properly calibrated to the Sun \citep{Gough1976,Pedersen1990,Salaris2008}. In
\mes the formulation from \citet{Cox1968} is employed, while \gar relies on the
prescription from \citet{Kippenhahn2012}.

As mentioned earlier, the simulations use a non-standard solar mixture. This
mixture will be used in \gar, while we for the \mes models settle for the almost
identical GN93.

The atmospheric simulations use the Mihalas-Hummer-D\"{a}ppen (MHD) equation of
state (EoS) \citep{Hummer1988,Mihalas1988,Dappen1988}, which we have readily
available in \gar, but not in \mes. To span the temperature range necessary for
a full structure model, we complement the MHD-EoS with the OPAL-EoS
\citep{Rogers2002} in \gar. In \mes we only use OPAL-EoS.

\citet{Trampedach2014a} calculated their own low-temperature opacities for the
3D simulations. In the envelope model used for the \mlt-calibration, these were
merged with interior opacities from the Opacity Project
\citep[OP,][]{Badnell2005}. We have used the available OP data to calculate
opacity tables for the quite specific mixture used in the 3D simulations and
merged them with the atmospheric opacities from \citet{Trampedach2014a} provided
by R.~Trampedach (priv. comm.). These custom opacities are used in \gar. In
MESA, we instead combine the low-temperature Trampedach-opacities with tables
from OPAL \citep{Iglesias1996}. For completeness it should be added that we use
the Potekhin conductive opacities \citep{Cassisi2007}.

\subsection{Solar calibration}
\label{sec:results_suncal}

The first step is to calculate a standard solar model (SSM) using our new
implementation. This is done by performing a solar calibration, which determines
the initial chemical composition and the \mlt scaling factor from \eref{eq:mlt}.
Microscopic diffusion -- which is required to perform the procedure -- is
included in all of our models.

A solar calibration with the 3D results activated yields a scaling factor of
$\kmlt = 1.024$ for \gar and $\kmlt = 1.034$ for \mes. The value of \mlt for the
solar simulation in the grid of 3D simulations is $\mltgs = 1.764$; thus, the
corresponding values are $\mlts = 1.807$ for \gar and $\mlts = 1.828$ for \mes.

In the following we will compare the models with our new 3D implementation to
\emph{standard reference} models, i.e., evolution with a constant,
solar-calibrated \mlt and Eddington grey atmosphere, but otherwise using the
same microphysics. When performing these standard Eddington solar calibrations
we obtain virtually the same initial abundances as with our 3D implementations.
For reference, we get $\mlts = 1.702$ for \gar and $\mlts = 1.672$ for \mes,
which is already a clear indication that the outermost structure has changed
compared to the 3D case.

\subsection{Evolution}
\label{sec:results_evol}

\begin{figure}
  \centering
  \includegraphics[width=0.98\columnwidth]{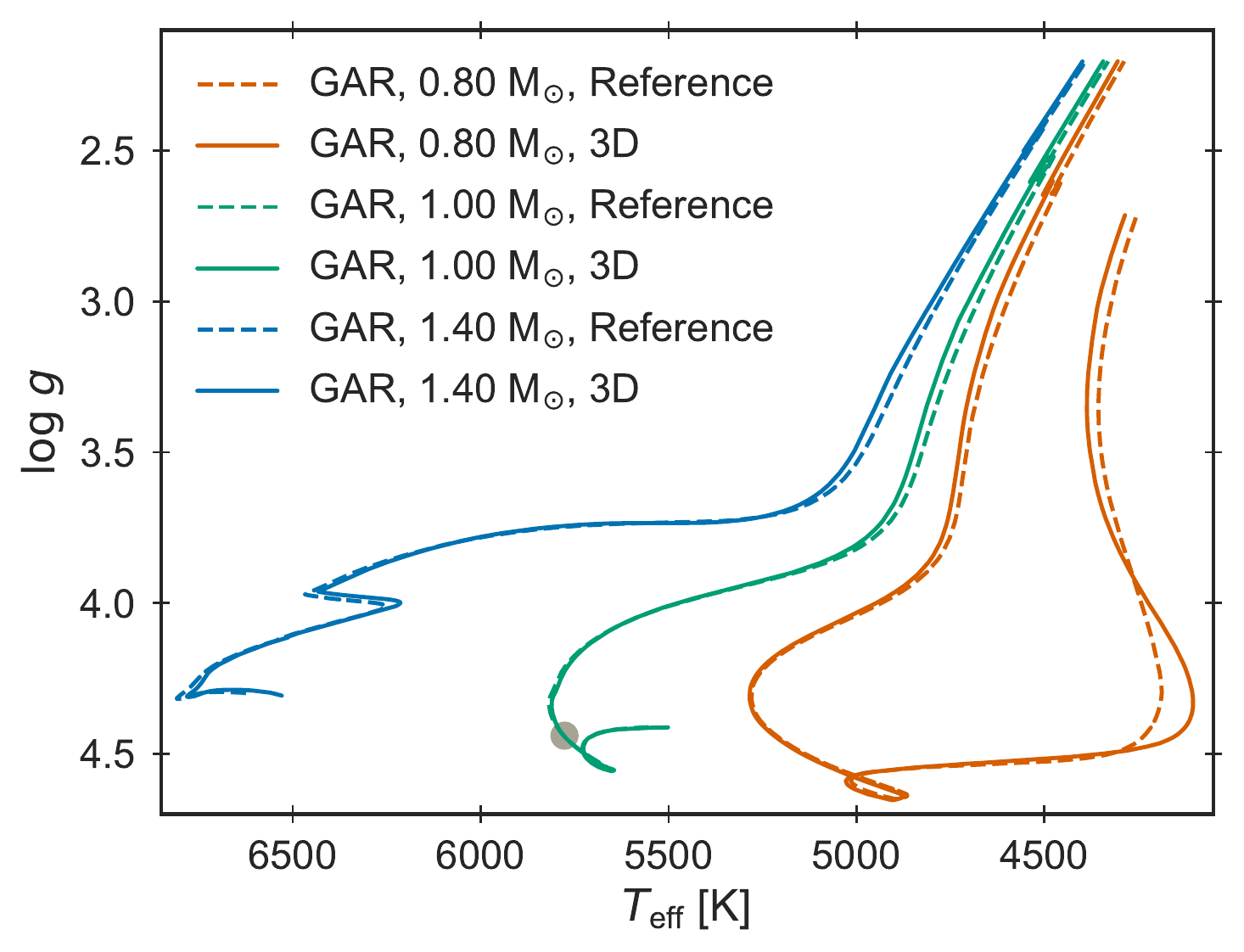}
  \caption{Stellar evolutionary tracks from \gar. The \emph{reference}-tracks
    are computed using a constant, solar calibrated \mlt and an Eddington
    atmosphere. The \emph{3D}-tracks use our implementation of variable \mlt and
    a \ttaurel{} from 3D simulations. To not clutter the plot, the full pre-main
    sequence phase is only shown for the $0.80\msun$ track. The present-day Sun
    is marked in grey.}
  \label{fig:tracks_all}
\end{figure}

\begin{figure}
  \centering
  \includegraphics[width=0.98\columnwidth]{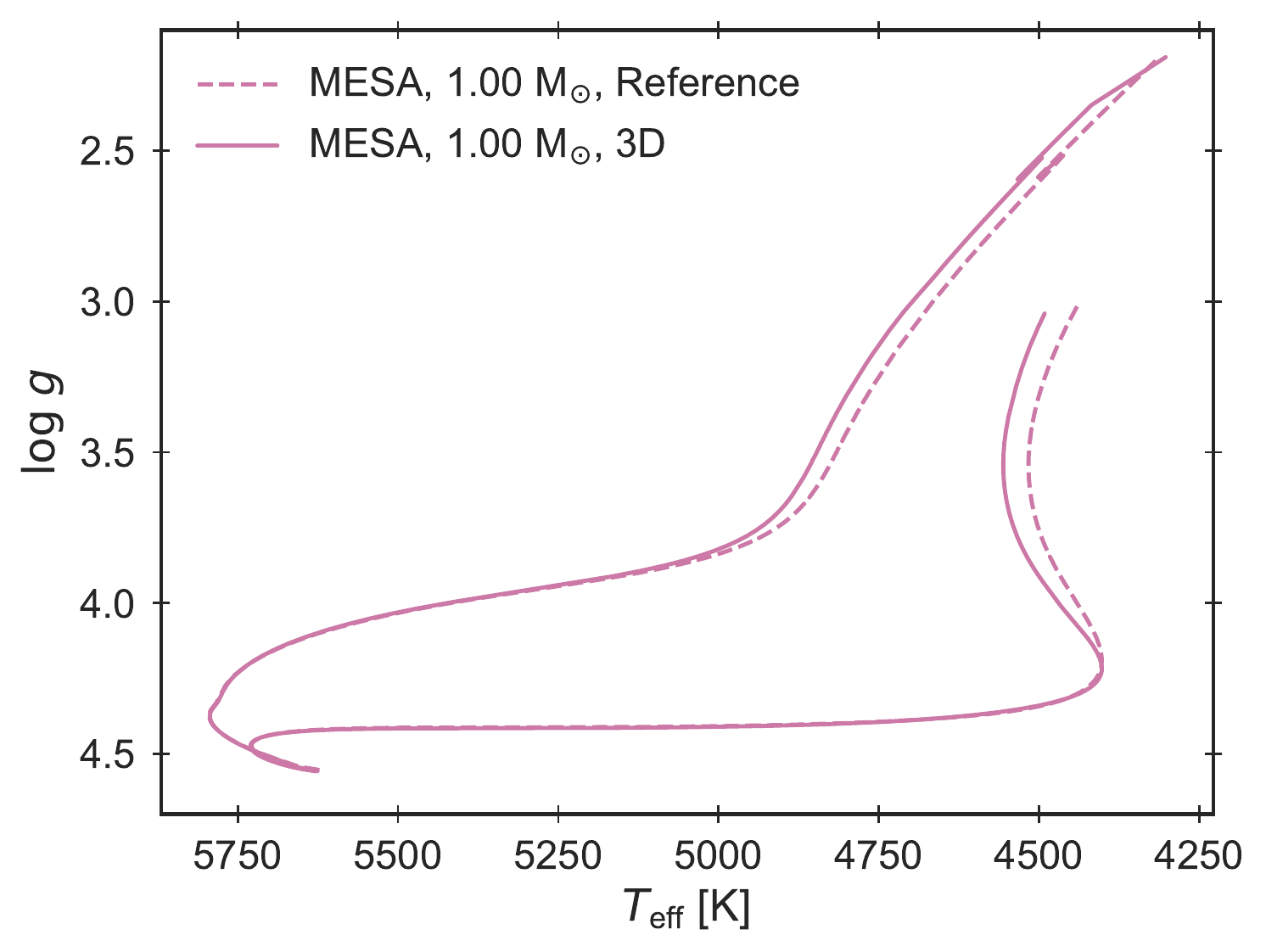}
  \caption{As Figure~\ref{fig:tracks_all}, but calculated with \mes and showing
    only the $1.00\msun$ track including the PMS phase. The extrapolation in the
    RGB is treated differently (details in the text).}
  \label{fig:track_1msun}
\end{figure}

After the solar calibration, the first natural test of our new implementation is
to calculate evolutionary tracks and compare to reference tracks. This is shown
in \fref{fig:tracks_all} and \fref{fig:track_1msun}.
%
%

From the plots it is clear that the main sequence of the $1.00\msun$ evolution
is nearly identical for the two tracks. This is just as expected, as the solar
calibration ensures that both of them go through the same point at the Sun's age
(highlighted in grey in the figure). The turn-off is almost identical as well,
with a temperature difference of less than $7\kel$. The same is the case for the
turn-off of the $0.80\msun$ track, where the tracks differ by $5\kel$. For the
$1.40\msun$ evolution, the tracks are separated by $35\kel$ at the leftmost
point in the hook.

The two sets of tracks are more clearly distinguished from each other on the
pre-main sequence (PMS) and on the red-giant branch (RGB), where convection, and
in particular its super-adiabatic parts, are most extended. For the \gar tracks,
the temperature difference in the RGB at $\logg = 3.2$ between \emph{3D} and
\emph{reference} is $25\kel$ for the $0.80\msun$ track, $28\kel$ for
$1.00\msun$, and $29\kel$ for $1.40\msun$. At the RGB-bump, the difference is
unchanged in the case of $0.80\msun$, roughly halved for $1.00\msun$, and
reduced to $11\kel$ for the $1.40\msun$ track. The RGB-bump for the $1.00\msun$
\mes tracks differ by $32\kel$.

As expected, the evolutionary pace of the models is also almost unchanged by our
new implementation. At the exhaustion of hydrogen in the core (defined as a
central hydrogen content of less than $10^{-5}$), the age difference between the
two sets of tracks is less than $0.1\%$ for all of the masses. The maximum age
difference for the $1.00\msun$ evolution is obtained, if we instead define the
turn-off in a more observational sense -- i.e., where \teff reaches its highest
value -- which yields a change in age of around $2\%$ (or $150\,\text{Myr}$).

From the work by \citet{Salaris2015}, the \ttaurel is expected to play the
largest role in the temperature change (see also \sref{sec:discussion}). We find
that the variable \mlt plays an important role as well -- especially in the RGB.
A fixed \mlt makes the two tracks move up the RGB in parallel, i.e., with a
constant separation. However, with the 3D implementation, the variations in \mlt
as the star evolves will give rise to changes in the slope of the ascent. This
is visible from the varying temperature differences along the RGB evolution.

\subsubsection{Limitations in the RGB}
\label{sec:results_evol_limits}

One clear limitation of the method is the coverage of the grid, and therefore
the treatment of the borders is important. If the 3D-implementation is fully
switched off when the star leaves the valid grid range, the track will suddenly
experience a jump in \mlt as well as a different outer boundary condition, and
changes temperature accordingly. To avoid this, we rely on extrapolation
just outside the boundaries of the grid, which is especially important in
the fully convective PMS phase, where the lower mass tracks will leave the grid
very briefly. We have tested this near the boundaries at high \logg for
both the hot and cold edge -- i.e. for stars with higher/lower mass than the
$0.80$-$1.40\msun$ range -- where the transition occurs smoothly.

On the RGB -- where extrapolation is required to continue the evolution towards
lower \logg -- the situation is a bit more complicated. From the top right
corner of \fref{fig:track_1msun} it can be seen that the 3D-track suddenly cools
and crosses the reference track. What is happening is that the track leaves
the grid resulting in a fast drop in the value of \mlt, which effectively
makes the track change to a different adiabat and continue its ascent.
Clearly, simple extrapolation from the triangulation does not work properly
in this region. What is done instead in \gar is to use the last valid values
from the grid -- i.e. keeping \mlt and the \ttaurel fixed during RGB evolution.
This produces a smooth evolution as can be seen from \fref{fig:tracks_all}
, but is not to be trusted going high up the RGB.

\subsection{Model structure}
\label{sec:results_struc}

We now compare the structure of the calibrated solar models with the 3D \mlt and
\ttaurel, to the reference standard solar models with an Eddington \ttaurel and
constant solar calibrated \mlt. As the implementation only affects the layers
above and just below the photosphere, this is where we mainly expect to see
changes in the models.

\begin{figure}
  \centering
  \includegraphics[width=0.98\columnwidth]{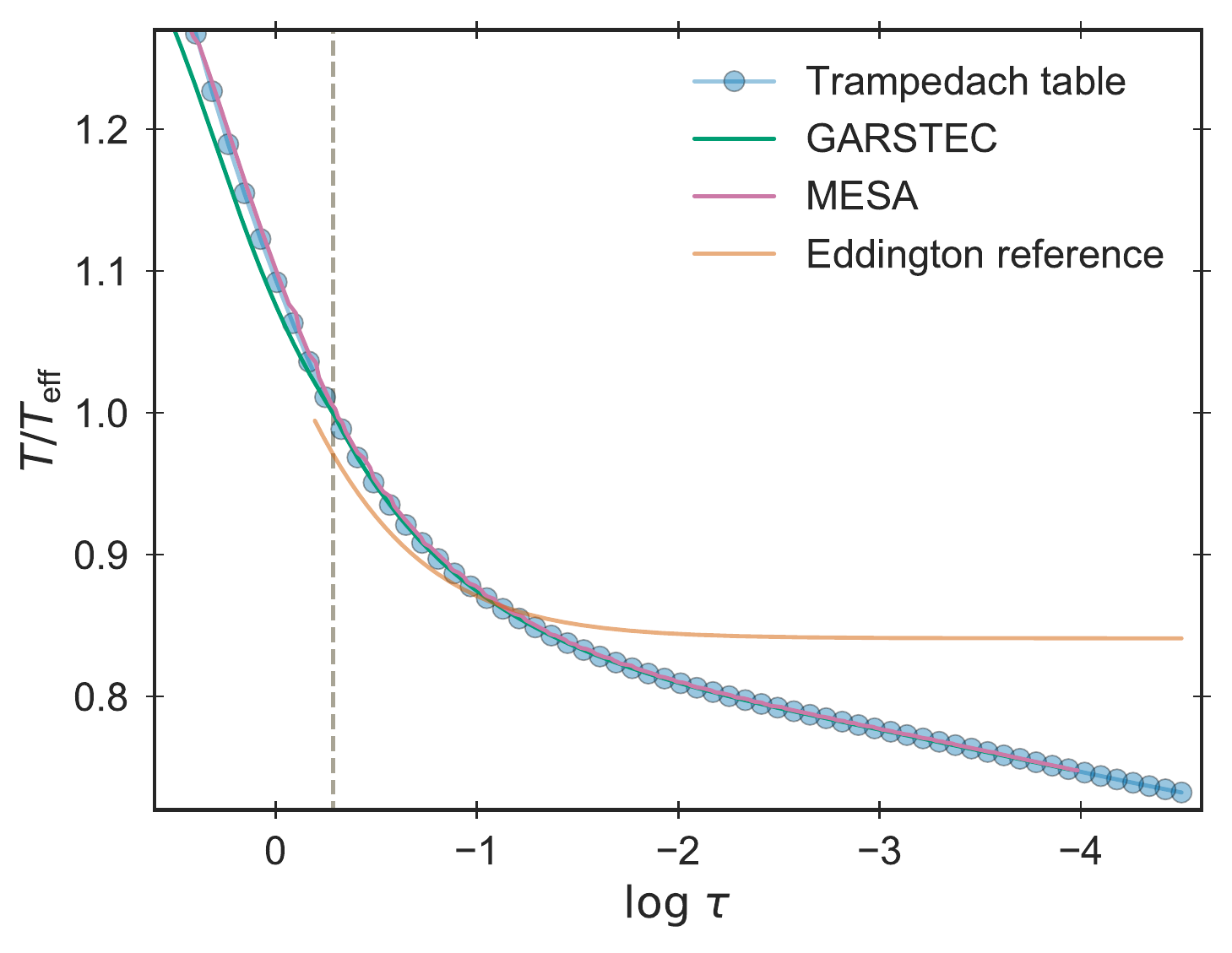}
  \caption{Temperature (normalised to the value at the photosphere) as a
    function of optical depth in the outermost parts of the \gar and \mes 3D
    solar models. The grey dashed line marks the photosphere of the stellar
    structure models. The curve \emph{Trampedach table} shows the 3D \ttaurel of
    the solar simulation from the table. Finally, the result using an Eddington
    grey \ttaurel is added for reference.}
  \label{fig:ttau_compare}
\end{figure}

In \fref{fig:ttau_compare}, the temperature structure in the form of \ttau is
displayed for the solar models with 3D effects activated from both \mes and
\gar. They are compared to the \ttaurel from the solar entry in the table from
\citet{Trampedach2014a,Trampedach2014b}. The transition point at $\log\tau
\simeq -0.3$ (corresponding to $\tau \simeq 0.5$) is marked for clarity. For
reference, the Eddington grey \ttaurel -- which is anchored at $\tau = 2/3$ and
therefore differs at the marked photosphere -- is also displayed in the figure.
The figure clearly shows that both models trace the 3D \ttaurel used in the
calculation closely in the atmosphere.
%

\begin{figure}
  \centering
  \includegraphics[width=0.98\columnwidth]{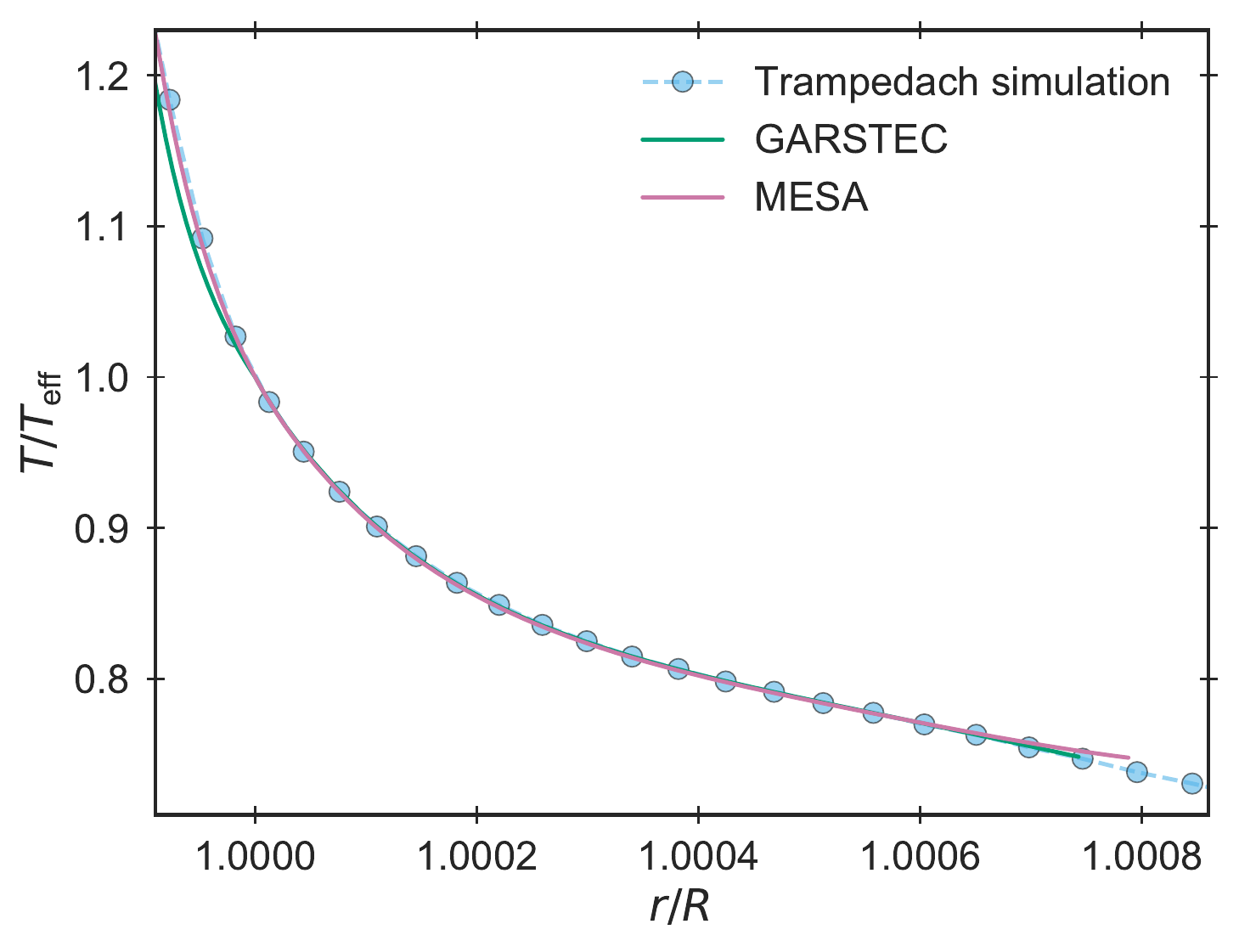}
  \caption{As \fref{fig:ttau_compare} but shown as a function of radius
    (normalised to the photospheric value). The \emph{Trampedach simulation}
    is the temperature stratification from the averaged 3D simulation of the
    Sun from the grid.}
  \label{fig:trad_compare}
\end{figure}

We show the temperature stratification as $T(r)$ for the same solar models in
\fref{fig:trad_compare}. In this figure, they are compared to the actual
structure extracted directly from the 3D simulation of the Sun (provided by
R.~Trampedach, priv. comm.)\footnote{Thus, the reference curves in
  \fref{fig:ttau_compare} and \fref{fig:trad_compare} are different: The former
  is from the table of extracted results, the latter directly from the averaged
  simulation. This also explains the different sampling of the reference curves
  in the two figures.}. It is evident that the models very accurately reproduce
the full underlying averaged 3D simulation. Thus, the stellar models are able to
recover the actual temperature stratification from the atmosphere simulation.
Below the photosphere the models naturally deviate as convection starts to
influence the structure, which is also the case in \fref{fig:ttau_compare}.
%

\subsubsection{Stellar interior}
\label{sec:results_struc_interior}

Turning to the interior of the stellar models, we see that the implementation of
the 3D results has an insignificant impact on the structure. In general the
models are indistinguishable -- e.g. with respect to the hydrogen profile.
Specifically, the depth of the outer convection zone is left virtually
unchanged; the relative difference between 3D and reference is below $0.01\%$
for both \mes and \gar.



\subsubsection{Oscillation frequencies}
\label{sec:results_struc_osc}
Asteroseismology is an excellent tool for probing the interior of stars, by
observing the imprint of the stellar oscillations in the emitted light. The term
\emph{solar-like oscillations} governs stochastically excited waves in stars
with an outer convection zone -- i.e. stars on the lower main sequence and red
giants. The advent of very high quality space-based photometry (from e.g.
\emph{Kepler}) has made it possible to detect such oscillations in distant stars
allowing us to gain detailed knowledge of their internal structure \citep[e.g.
the review by][]{Chaplin2013}. For stars showing solar-like oscillations, it has
revolutionised how well we can determine stellar properties
\citep{Lund2017,Aguirre2017}.

\begin{figure}
  \centering
  \includegraphics[width=0.98\columnwidth]{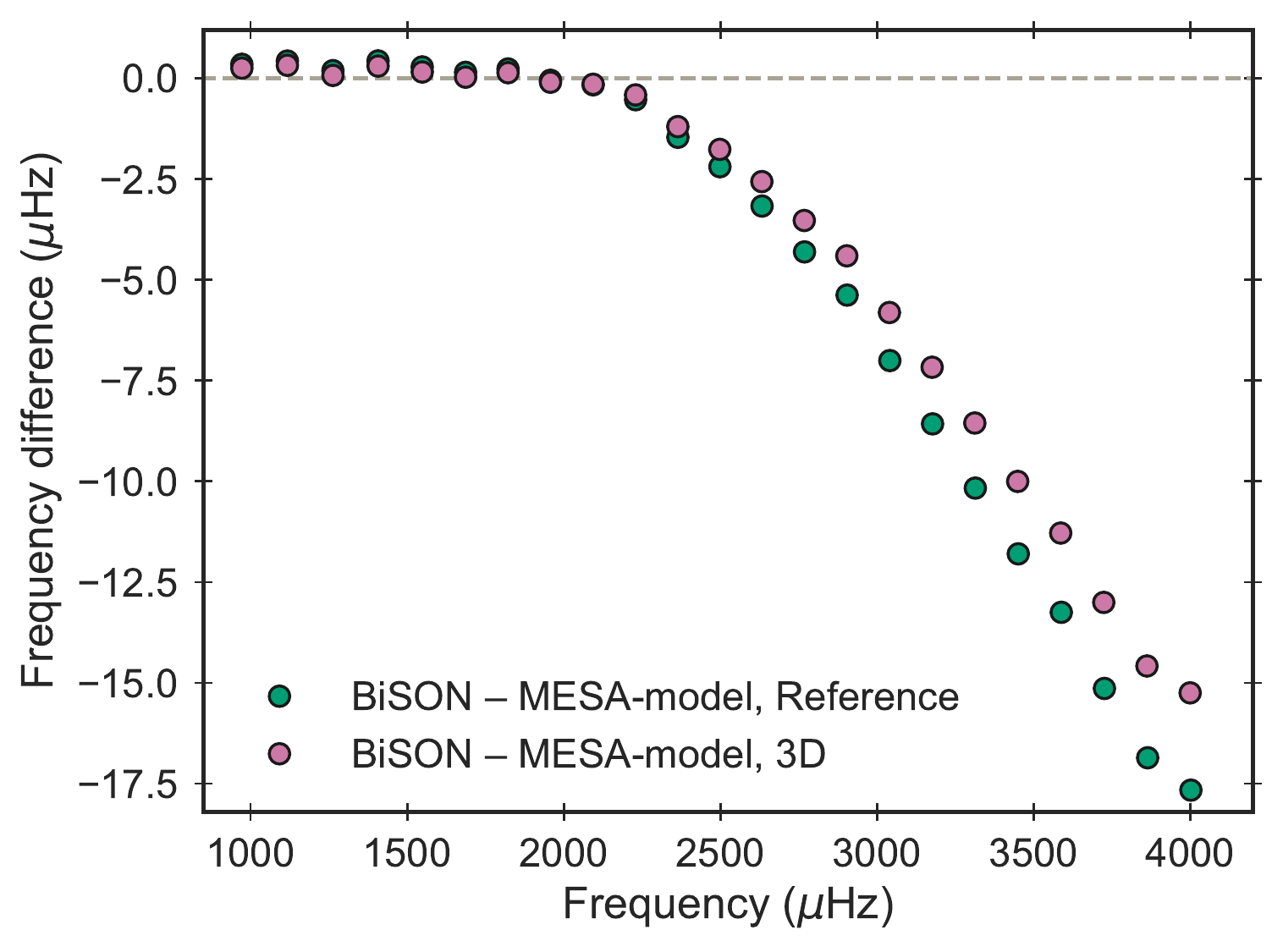}
  \caption{Difference between observed and calculated oscillation frequencies of
    the Sun. The observations are from BiSON and the model frequencies are from
    the \mes solar models with and without our modifications.}
  \label{fig:osc_mesa}
\end{figure}

In the present case, we can compare the structure of the calibrated solar models
with the observed frequencies from the Sun. We have calculated theoretical
oscillation frequencies for our solar models using the Aarhus adiabatic
oscillation package \citep[\adi,][]{Christensen-Dalsgaard2008a}. As observations
we use solar data from the Birmingham Solar-Oscillation Network
\citep[BiSON,][]{Broomhall2009,Davies2014}. The comparison is shown in
\fref{fig:osc_mesa} as a difference between the two set of oscillation
frequencies for the \mes solar models.
%

The general deviation, which increases for higher frequencies, is well-known and
expected: it is the asteroseismic surface effect (mentioned in
\sref{sec:atmos_patch}), consequence of the near-surface deficiencies in stellar
models. Oscillations of higher and higher frequency probe regions closer and
closer to the surface; thus, it is evident from the figure that the two models
differ in the surface layers. This is just as anticipated, as the implementation
changes the outer boundary condition and the correction from \eref{eq:radgrad}
is only applied just below the photosphere (because the optical depth increases
very rapidly in the interior). Hence, the inclusion of the 3D effects shifts the
oscillation frequencies, which is a known effect of changing the atmospheric
structure \citep[e.g.][]{Morel1994}. The frequencies are decreased, thus
seemingly bringing the model closer to the observations.

The surface effect can be further improved on by fully replacing the outer
layers of the model with an averaged 3D simulation in a so-called patched model
(see \sref{sec:atmos_patch} and references therein). However, this is
(currently) only performed as a final step after the stellar evolution
calculation; it is not possible to perform the procedure along the way during
the evolution. Furthermore, a full treatment of the surface effect would
additionally involve the inclusion of the \textit{modal} effects, i.e. the
effects of nonadiabaticity and the full interaction between convection and
pulsations \citep[e.g.][]{Houdek2017}, as is also evident from the results of
patched models \citep[e.g][]{Rosenthal1999,Sonoi2015,Ball2016,Joergensen2017}.

\section{Comparison to earlier work}
\label{sec:discussion}

An analysis similar to the one we present in this paper was carried out by
\citet[][hereafter SC15]{Salaris2015} using the stellar evolution code BaSTI
\citep[BAg of Stellar Tracks and Isochrones,][]{Pietrinferni2004}. They
implemented the same 3D results, namely the calibrated \mlt and \ttaurel{}s from
\citet{Trampedach2014a,Trampedach2014b}. However, the description of their
implementation is brief and differs from ours in a few ways.

One central topic is the transition point between atmosphere and stellar
interior, and the location of the photosphere in the model. SC15 used a constant
$\taufit=2/3$ as transition point (and mention that they have tested various
other values down to $\taufit=100$). The crucial matter is not the exact value,
but whether this transition point has been used as the photosphere in the
stellar model or not. The optical depth $\taueff$ in the 3D \ttaurel at which $T
= \teff$ varies, but is always close to $\taueff \simeq 0.5$. As argued earlier
in \sref{sec:implement_ttau_atmos}, it will produce a slightly inconsistent
model, if the global `photospheric quantities' (e.g. \teff and \logg) are still
determined at the usual $\taufit = 2/3$ or another
$\taufit\neq\taueff$.\footnote{Especially if these quantities are passed to the
  3D interpolation routine.} We have taken great care in this respect, either by
making sure the transition happens at the correct $\taufit = \taueff$ each time
(\gar) or by properly determining the global quantities at this point (\mes).

SC15 mention the use of \citet[][eq. (35), (36)]{Trampedach2014a} to correct the
temperature gradients. First of all, as explained earlier, it is only necessary
to apply a correction to the radiative temperature gradient \radgrad (by using
eq.~(35) from \citet{Trampedach2014a}); the MLT calculation with this modified
gradient as input will ensure a properly corrected $\nabla$ without further
modifications (R.~Trampedach, priv. comm.). Secondly, SC15 find the corrections
to be minuscule \footnote{SC15 state a difference between the two sets of
  temperature gradients of `much less than $1\%$'.}. We agree for higher values
of $\tau$, but just below the photosphere the correction term in
\eref{eq:radgrad} is significant -- e.g. for the solar models, the temperature
gradient is changed by around $20\%$ at $\tau \simeq 1$. A plot of the
correction is shown in Appendix~\ref{app:gradcorr}.

As argued earlier -- and also by \citet{Ludwig1999} and \citet{Trampedach2014b}
-- it is appropriate to introduce a calibrated scaling factor instead of using
\mlt directly as supplied from the grid. SC15 mention such a scaling as required
to produce a standard solar model (see below). However, they did not employ this
\mlt scaling factor in their stellar evolution calculations.

\subsection{Results}
\label{sec:discussion_results}

A general difference between our work and SC15 is the focus of the
investigation. SC15 did calculate models utilising both the 3D \mlt and
\ttaurel, but also analysed them separately in order to isolate the impact of
the 3D \mlt and \ttaurel, respectively. We choose to follow the advice of the
original work by \citet{Trampedach2014b}, who stress the importance of
\emph{always} employing the extracted results together (as well as using them
alongside the corresponding atmospheric opacities).
%

For their comparison tracks (with different \ttaurel{}s), SC15 used the \mlts
from the 3D RHD solar simulation directly, instead of a traditional solar
calibrated value. We have chosen to make solar calibrations for both the 3D and
reference Eddington case separately, in order to show the difference between
using the 3D results and what ``modellers usually do''. Thus, a direct
comparison of our results to SC15 is not possible. For example, in SC15 the
Eddington tracks are generally hotter than the 3D ones while we see the opposite
behaviour. Looking at figure 3 in SC15, this is an effect of the choice to not
individually calibrate the reference tracks to the Sun in SC15; the
main-sequence evolution of the $1.00\msun$ Eddington track is significantly
hotter than the 3D counterpart, where our tracks basically coincide. If we
perform evolutionary calculations with similar assumptions -- i.e. disable
scaling of \mlt for the 3D case, and for the reference case directly use the
\mlts from the grid instead of a solar-calibrated value -- we obtain results
which are very similar to SC15 (see Appendix~\ref{app:sc15}).

SC15 briefly analyse the impact on the standard solar model (SSM) and state that
they ``find it necessary to rescale the RHD \mlt calibration by a factor of just
$1.034$ to reproduce the solar radius''. This is exactly the same we find for
the \mes 3D solar calibration and very close to our \gar value. However, this
scaling factor was not used to calculate their evolutionary tracks. They do not
dig deeper into the interior structure of the SSM, so we are unable to compare
our asteroseismic analysis to their work.



\section{Conclusions}
\label{sec:conclusion}

We have consistently implemented results extracted from 3D radiation-coupled
hydrodynamics simulations of stellar convection in the stellar evolution codes
\gar and \mes. The new implementation consists of a temperature stratification
in the form of \ttaurel{}s and corresponding corrected temperature gradients,
and a calibrated, variable mixing-length parameter $\mlt(\teff,\logg)$. We have
presented a very detailed account of our implementations, and compared to the
earlier implementation of the same 3D results in a stellar evolution code by
\citet{Salaris2015}. Moreover we make our \mes implementation freely available
(see the appendix).

We calculate the evolution of different low-mass stars after a solar
calibration. We compare to a set of reference models which uses an Eddington
grey atmosphere and constant solar-calibrated \mlt. In the pre-main sequence and
in the red-giant branch we see the largest effect of the 3D implementation on
the temperature evolution. Regarding the evolutionary speed, we see no
significant change in the age of the models at core hydrogen exhaustion.

Furthermore, we compare the model structure of the calibrated solar models and
see no significant changes. For the first time we present an helioseismic
analysis of a standard solar model with a \ttaurel and variable \mlt extracted
from 3D simulations. The use of the 3D effects makes a small, positive impact on
the asteroseismic surface effect.

We note that the method is limited by the coverage in parameter space of the
grid of 3D simulations. To make this method widely applicable, an extended
coverage in terms of \logg and \teff is required, but more importantly:
simulations of varying metallicity. One option in this respect would be to
extract similar information from the \textsc{stagger}-grid by \citet{Magic2013}.

The future aim is to exploit the concept of patched models -- which provide a
more realistic structure but is static -- with the current dynamic approach, to
allow for the evolution of more realistic models. Our implementation is one step
further along that path and the work by \citet{Joergensen2017} another. The
ultimate goal is to facilitate on-the-fly patching of 3D simulations as outer
boundary conditions for stellar evolution calculations.

\section*{Acknowledgements}

We thank the anonymous referee for the review and suggestions. We
gratefully thank R.~Trampedach for making his interpolation routine and
opacities available, for opinions on technical matters and fruitful discussions.
We also wish to express our thanks to A.C.S.~J{\o}rgensen for collaboration on
the 3D simulations and technical matters in \gar. Finally, we thank S.~Cassisi
for clarifications regarding the work in SC15 and for many useful
comments. Funding for the Stellar Astrophysics Centre is provided by The
Danish National Research Foundation (Grant DNRF106). WHB acknowledges funding
from the UK Science and Technology Facilities Council (STFC). VSA acknowledges
support from VILLUM FONDEN (research grant 10118).




\bibliographystyle{mnras}
\bibliography{manual_refs,mendeley_export}





\appendix

\section{Technical details of the \mes implementation}
\label{app:mesa}

The opacity tables, \ttaurel and mixing-length parameters \mlt for the
simulations by \citet{Trampedach2013,Trampedach2014a,Trampedach2014b} are
available from the MESA
marketplace\footnote{\url{http://cococubed.asu.edu/mesa_market/}} as an archive
that should be extracted over an existing installation of \mes revision 9575.
Code to use the \ttaurel in \mes runs is provided in two examples in a new
folder \texttt{hydro\_Ttau\_examples}: \texttt{calibrate\_hydro\_Ttau} and
\texttt{evolve\_hydro\_Ttau}. The archive also includes code for the equivalent
`reference'' runs with an Eddington grey atmosphere. These are in the folders
\texttt{calibrate\_edd} and \texttt{evolve\_edd}.

The first example, \texttt{calibrate\_hydro\_Ttau}, provides an example case
that was used to calibrate the scaling factor presented in
Sec.~\ref{sec:results_suncal}. It implements a solar calibration using the
\texttt{astero} module, with a small Python script that optimizes the model
parameters. This example can be adapted by users wishing to use the 3D
\ttaurel{} alongside the \texttt{astero} module. The second example,
\texttt{evolve\_hydro\_Ttau}, provides an example used to compute the
evolutionary track presented in Sec.~\ref{sec:results_evol}, and can be adapted
for normal evolutionary calculations. In both examples, the code that
interpolates the 3D \ttaurel and MLT parameters $\mlt$ and modifies the boundary
conditions and radiative gradient is included in the files
\texttt{run\_star\_extras.f} and \texttt{624.dek}. The code for these two
examples is also now part of the main \mes codebase and will appear in the test
suite of \mes next public release (after r10398). Because of small changes
elsewhere in \mes, the results of those test cases are not identical to the
results presented here.

\subsection{Modification of radiative gradient}
\label{app:rad_grad}

Since revision 9575, MESA has included a ``porosity'' factor $\phi$ that reduces
the opacity. By default, when $\phi>1$, the opacity $\kappa$ is replaced by
$\kappa/\phi$ when computing the radiative gradient, so the radiative gradient
$\radgrad$ is replaced by $\radgrad/\phi$. To modify the radiative gradient as
in eq.~(\ref{eq:radgrad}), we therefore assign
\begin{equation}
  \phi=\frac{1}{1+q'(\tau)}\;,
\end{equation}
where $q(\tau)$ has been extracted as described in Sec.~\ref{sec:implement}. It
is important to note that the changes to the porosity factor do \emph{not}
currently affect the optical depth $\tau$. If this should change in subsequent
versions of MESA, the current implementation will no longer be valid.

In the main codebase for \mes r9575, the porosity factor is hardcoded to have a
minimum value of unity. In the $q(\tau)$ given by \citet{Trampedach2014a}, this
would limit us to $q'(\tau)<0$, so our archive includes modified versions of
core source files (\texttt{hydro\_eqns.f90} and \texttt{mlt\_info.f90} in
\texttt{star/private}) that lift this restriction on the porosity factor.
The restriction is also lifted in public releases after r10398 by
the creation of a user control for the porosity limit.

\subsection{Boundary conditions}
\label{app:mesa_boundary}
With the outermost meshpoint located above the photosphere, \mes needs new outer
boundary conditions. We choose these to correspond to an Eddington grey
atmosphere evaluated at the optical depth of the outermost meshpoint. For the
gas pressure $P_g$, we therefore have
\begin{align}
	P_g
    &=\left(\frac{g}{\kappa}-\frac{F}{c}\right)\tau \nonumber \\
    &=\left(\frac{g}{\kappa}-\frac{L}{4\pi R^2c}\right)\tau \nonumber \\
    &=\left(\frac{g}{\kappa}-\frac{Lg}{4\pi GMc}\right)\tau \; ,
\end{align}
where $g$ is the gravity, $\kappa$ the Rosseland mean opacity, $F$ the radiative
flux, $c$ the speed of light, $\tau$ the optical depth, $R$ the radius at the
outermost point, $G$ the gravitational constant and $M$ the total mass of the
star.

For the radiation pressure $P_r$, we have
\begin{align}
  P_r
  &=\frac{1}{3}aT^4 \nonumber \\
  &=\frac{4\sigma}{3c}\cdot\frac{3}{4}\teff^4(\tau+q(\tau)) \nonumber \\
  &=\frac{L}{4\pi R^2c}(\tau+q(\tau)) \nonumber \\
  &=\frac{Lg}{4\pi GMc}(\tau+q(\tau)) \; ,
\end{align}
where $a$ is the radiation constant, $T$ the temperature, $\sigma$ the
Stefan-Boltzmann constant, \teff the effective temperature and $q$ the Hopf
function (see eq. \ref{eq:generalhopf}).

The total pressure $P$ at the outermost meshpoint is therefore
\begin{align}
P(\tau)=P_g(\tau)+P_r(\tau)
&=\frac{g}{\kappa}\tau+\frac{Lg}{4\pi GMc}q(\tau) \nonumber \\
&=\frac{g}{\kappa}{\tau}\left[1+\frac{\kappa}{\tau}
  \frac{L}{4\pi GMc}q(\tau)\right] \; .
\end{align}
These are, in effect, equivalent to the initial conditions that \mes uses when
integrating an atmosphere from small optical depths down to the photosphere to
obtain photospheric boundary conditions.

\section{Models with SC15 settings}
\label{app:sc15}

\begin{figure}
  \centering
  \includegraphics[width=0.98\columnwidth]{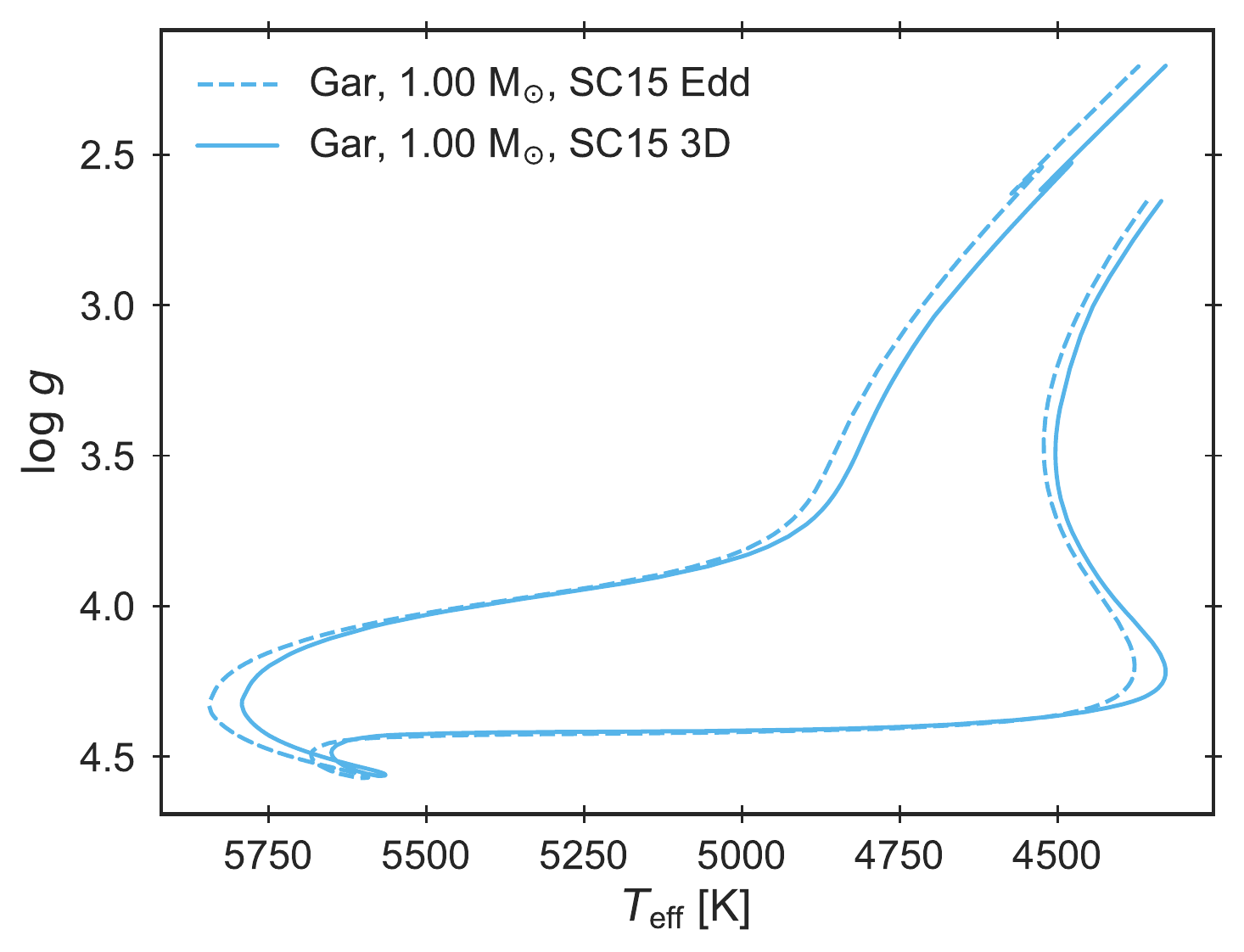}
  \caption{Stellar evolution calculated with \gar using same settings as SC15
    (see text for details).}
  \label{fig:track_sc15}
\end{figure}

Here we present models calculated with \gar utilising similar settings as those
used by SC15 \citep[i.e.][]{Salaris2015}. The 3D case is calculated without
scaling of \mlt, i.e. just by adopting the values directly from the grid. The
reference Eddington tracks do not rely on a solar calibrated \mlt, but rather
the value \mlts directly from the solar 3D simulation. Moreover, the tracks are
calculated with the GN93 composition and without diffusion. The resulting tracks
for the $1\msun$ star can be seen in \fref{fig:track_sc15}, which is very
similar to the corresponding track in figure 3 of SC15. The figure can also be
compared to our models in \fref{fig:tracks_all} and \fref{fig:track_1msun}.
%

\section{Correction of the gradients}
\label{app:gradcorr}

\begin{figure}
  \centering
  \includegraphics[width=0.98\columnwidth]{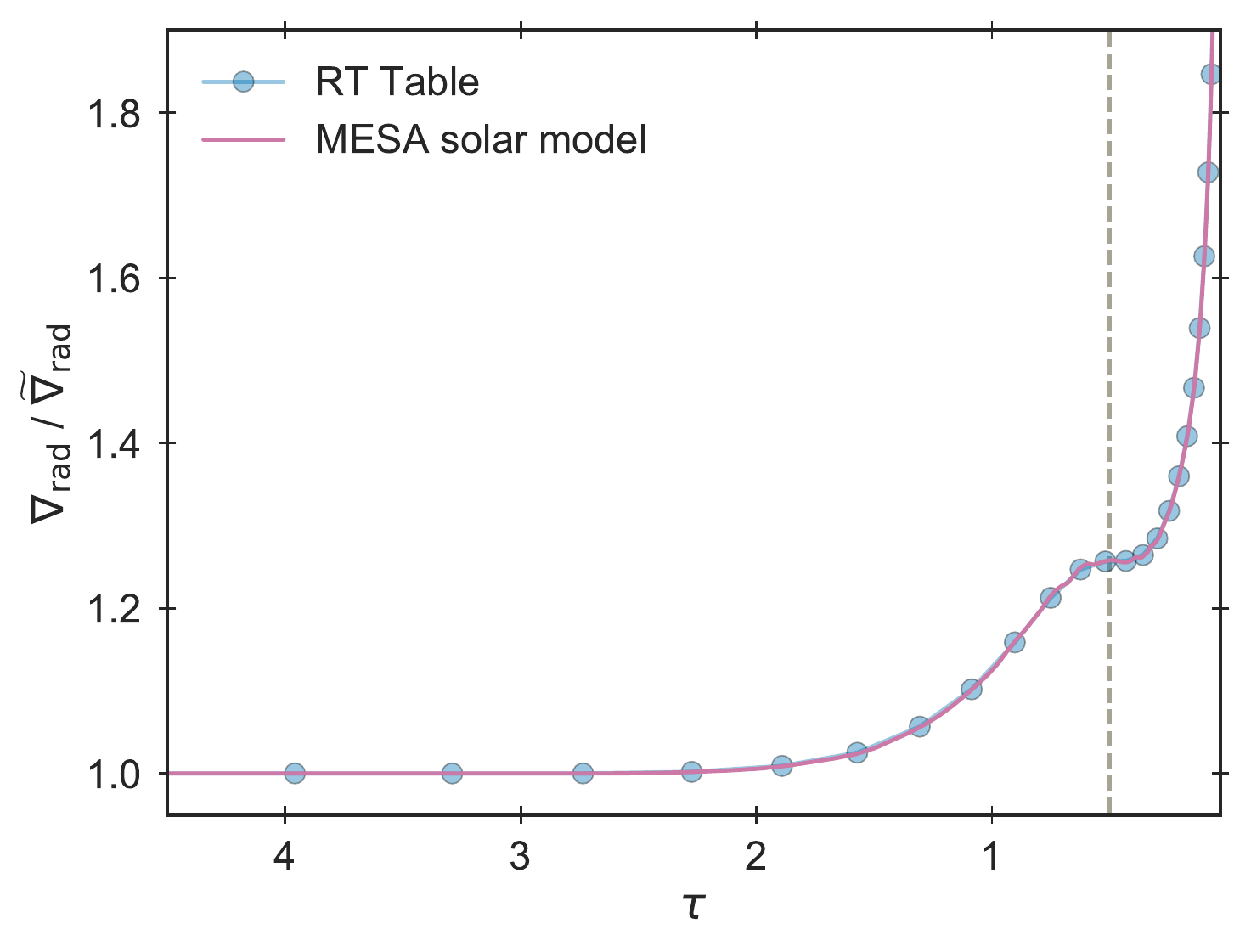}
  \caption{Magnitude of the correction term in \eref{eq:radgrad} for the
    \mes solar model and derived directly from the table of 3D results. The
    grey dashed line marks the photosphere at $\taueff\simeq0.5$.}
  \label{fig:gradcorr}
\end{figure}

In this appendix we highlight the impact of the correction factor in
\eref{eq:radgrad} on the temperature gradients. The result from the \mes solar
model as well as $1 + \dqq$ derived directly from the Trampedach data table is
shown in \fref{fig:gradcorr} in the region near the photosphere. It is clear
that the changes are significant around the photosphere and quickly decrease
going into the star -- and that the model accurately reproduce the table.
%

\bsp	
\label{lastpage}
\end{document}